# P-Odd Asymmetry in the Deuteron Disintegration by Circularly Polarized Photons

I.B. Khriplovich[1] and R.V. Korkin
Budker Institute of Nuclear Physics
630090 Novosibirsk, Russia,
and Novosibirsk University

**Abstract**

We calculate P-odd difference of the total cross-sections of the deuteron disintegration by left and right polarized photons. The relative magnitude of this difference in the threshold region is about $10^{-7}$. Its experimental measurement would give valuable information on the weak nucleon-nucleon interaction at short distances.

---

[1]khriplovich@inp.nsk.su

*To the memory of Volodya Telitsin,*
*our friend and enthusiastic advisor*
*on nuclear physics and physics in general*

# 1 Introduction

The deuteron, being the simplest nuclear system, in many cases allows for a relatively reliable theoretical analysis. That is why the problem of parity nonconservation (PNC) in the deuteron for a long time attracts attention of both experimentalists and theorists. Unfortunately, PNC effects in the deuteron are tiny, so that up to now only upper limits on them have been obtained experimentally [1–3].

At present, however, new prospects have arisen here due to creating intense sources of polarized photons, electrons, and neutrons. On the other hand, now the experimental investigations of PNC effects in the deuteron have become of great interest. One may hope that they will resolve a contradiction which exists at present in the problem of P-odd nuclear forces. The point is that recently the nuclear anapole moment (AM) of $^{133}$Cs was discovered, and measured with good accuracy in atomic experiment [4]. The result of this experiment is in a reasonable quantitative agreement with the theoretical predictions, starting with [5, 6], if the so-called "best values" [7] are chosen for the parameters of P-odd nuclear forces. However, the results of some nuclear experiments indicate that the P-odd $\pi NN$ constant $\bar{g}$ is much smaller than its "best value" (see, e.g., [8, 9]). An obvious possibility to reconcile the results of the atomic experiment and the nuclear ones, is to assume that the magnitude of another, short-distance contribution to P-odd nuclear forces is essentially larger than its "best value".

We discuss in the present paper the P-odd asymmetry in the deuteron disintegration by circularly polarized $\gamma$-quanta. As will be demonstrated below, the component of the weak interaction which conserves the total spin, does not contribute to the effect. In particular, the weak $\pi$-meson exchange, which is (relatively) long-distance one, is not operative here. The discussed asymmetry is due to the short-distance P-odd interaction. Unfortunately, there is a serious problem with the theoretical description of the short-range P-odd effects. They are commonly described by means of $\rho$- and $\omega$-exchanges. However, the range of these potentials, $1/m_{\rho,\omega} \sim 0.3$ fm, is much smaller than the proton mean-square radius, $<r_p^2>^{1/2} \sim 0.8$ fm. Therefore, all calculations of P-odd effects based on weak $\rho$-, $\omega$-potentials (as well as using $\rho$-, $\omega$-potentials for the description of strong interactions) have no sound theoretical grounds. Our treatment of the short-distance weak interactions in the deuteron differs from that adopted in previous papers, but it is no exception in this respect. Still, there is an observation indicating that with our procedure the magnitude of nuclear P-odd effects at least is not overestimated. The point is that this procedure was used previously in [6] to derive the constant of the effective P-odd contact interaction of the valence nucleon with the nuclear core. Thus calculated value of the anapole moment of $^{133}$Cs in [6] is close to (in fact, even somewhat lower than) its experimental value obtained recently in [4].

Theoretical studies of PNC effects in the deuteron were started in [10–14]. Papers [10, 11] concentrated on the P-odd asymmetry in $d(\gamma, n)p$ reaction caused by linearly polarized photons. For the photon energies of several MeV this asymmetry is very small as compared to that due to circularly polarized $\gamma$-quanta, the latter being of interest to us. A phenomenological treatment of PNC effects in the deuteron was adopted in [12]. Later it was supplemented in [13] with quantitative estimates made in the dispersion approach and in the pion exchange model.

PNC effects in $ed$-scattering were considered in [15], but for a very special kinematics only.

The general problem of parity nonconservation in $ed$-scattering was investigated in [16–19], with detailed numerical estimates in [16, 18]. However, in this process the effect of the nuclear parity violation is masked by the direct P-odd $ed$ interaction due to weak neutral currents.

After [12-14], P-odd effects in the deuteron disintegration by circularly polarized $\gamma$-quanta (and in the inverse reaction) were addressed in [16,20–24]. Though our results are in a qualitative agreement with most of previous ones (see below), we believe that the present independent investigation of the important and interesting problem is worth efforts.

## 2 Wave functions, transition matrix elements, and cross-sections

The deuteron ground state is $^3S_1$ (a small $^3D_1$ admixture to it will be neglected throughout the paper). In the zero-range approximation (ZRA) its wave function is

$$\psi_d = \sqrt{\frac{\kappa}{2\pi}} \frac{e^{-\kappa r}}{r}. \tag{1}$$

Here $\kappa = \sqrt{m_p \varepsilon}$, where $m_p$ is the proton mass, and $\varepsilon = 2.23$ MeV is the deuteron binding energy. To the same approximation, the $^3S_1$ and $^1S_0$ wave functions of the continuous spectrum, $\psi_{St}$ and $\psi_{Ss}$, respectively, are (see, for instance, [19])

$$\psi_{St,Ss} = \frac{\sin pr}{pr} - a_{t,s} \frac{e^{ipr}}{r}, \qquad a_{t,s} = \frac{\alpha_{t,s}}{1 + i p \, \alpha_{t,s}}. \tag{2}$$

Here $\alpha_t = 5.42$ fm and $\alpha_s = -23.7$ fm are the triplet and singlet scattering lengths, respectively. At last, in the spirit of the zero-range approximation, for the $P$ state of the continuum we will use the free wave function

$$\frac{1}{p} \frac{d}{dr} \frac{\sin pr}{pr} .$$

Near the threshold the photodisintegration cross-section is dominated by $M1$ transition. Since the radial wave function of the deuteron is orthogonal to that of the $^3S_1$ state of the continuous spectrum, the $M1$ transition goes into the $^1S_0$ state. The expressions for the $M1$ matrix element and total cross-section, as calculated in the same zero-range approximation, are well-known (see, for instance, [25]):

$$\langle ^1S_0|(\mu_p\sigma_p + \mu_n\sigma_n)_z|^3S_1\rangle = \frac{2\sqrt{2\pi\kappa}\,(\mu_p - \mu_n)\,(1 - \kappa\alpha_s)}{(\kappa^2 + p^2)\,(1 - ip\,\alpha_s)}; \tag{3}$$

$$\sigma_{M1} = \frac{2\pi\alpha}{3} \frac{\kappa p\,(\mu_p - \mu_n)^2 (1 - \kappa\alpha_s)^2}{m_p^2\,(\kappa^2 + p^2)\,(1 + p^2\alpha_s^2)}. \tag{4}$$

Here $p$ is the relative momentum of final nucleons, $\alpha = 1/137$, $\mu_p$ and $\mu_n$ are the proton and neutron magnetic moments, respectively.

For higher energies the cross-section is dominated by $E1$ transitions into the $^3P_{0,1,2}$ states. The corresponding matrix element and total cross-section are, respectively:

$$\langle ^3P|\mathbf{r}|^3S\rangle = -4\sqrt{\frac{2\pi\kappa}{1 - \kappa r_t}} \frac{\mathbf{p}}{(\kappa^2 + p^2)^2}; \tag{5}$$

$$\sigma_{E1} = \frac{8\pi\alpha}{3} \frac{\kappa p^3}{(1 - \kappa r_t)\,(\kappa^2 + p^2)^3}. \tag{6}$$

The origin of the factors $(1-\kappa r_t)^{-1/2}$ and $(1-\kappa r_t)^{-1}$ in (5) and (6), respectively, is as follows. Large distances dominate in the matrix element $\langle^3P|\mathbf{r}|^3S\rangle$. In this asymptotic region the naïve ZRA expression (1) for the deuteron wave function must be augmented by a correction factor $(1-\kappa r_t)^{-1/2}$ (see [25, 26]), which obviously results in factor $(1-\kappa r_t)^{-1}$ in $\sigma_{E1}$. Here $r_t = 1.76(1)$ fm is the effective radius of the triplet state.

On the other hand, wave functions (1) and (2) have incorrect behaviour for $r \to 0$. It is of no importance for the above formulae, since short distances are inessential for $\sigma_{E1}$, and even for $\sigma_{M1}$. However, the situation is different for matrix elements of the P-odd weak interaction for which short distances are crucial. Here we will use model wave functions with more realistic properties. For the deuteron we choose [27]

$$\psi_d = \sqrt{\frac{\kappa}{2\pi(1-\kappa r_t)}} \frac{e^{-\kappa r_1}}{r_1}, \qquad r < r_1;$$

$$\psi_d = \sqrt{\frac{\kappa}{2\pi(1-\kappa r_t)}} \frac{e^{-\kappa r}}{r}, \qquad r > r_1. \qquad (7)$$

This wave function has the correct asymptotics at $r \to \infty$ (see above), tends to a constant at $r \to 0$, and at least is continuous everywhere. The numerical value $r_1 = 1.60$ fm is chosen in such a way that the wave function is normalized correctly. In fact, it is quite natural that this value is close to that of the triplet effective radius $r_t$. Let us note that the unphysical cusp of $\psi_d$ at $r = r_1$ is harmless for our problem since this wave function will enter integrands only (see formulae below).

As to the $^1S_0$ wave of the continuous spectrum, the potential for the singlet state is rather shallow, and the effective radius $r_s = 2.73(3)$ fm is larger than the triplet one. Thus, the variation of the wave function in the internal region is even more important here. We choose the model wave function for this state as

$$\psi_{Ss} = A \frac{\sin\sqrt{p^2+p_0^2}\, r}{\sqrt{p^2+p_0^2}\, r}, \qquad r < r_s;$$

$$\psi_{Ss} = \frac{\sin pr}{pr} - a_s \frac{e^{ipr}}{r}, \qquad r > r_s. \qquad (8)$$

Requiring the continuity of the wave function and its first derivative at $r = r_s$, we obtain

$$p_0 r_s = 1.5\,;$$

$$A(p) = \frac{\sqrt{p^2+p_0^2}\, r_s}{\sin\sqrt{p^2+p_0^2}\, r_s} \frac{\sin pr_s - p\alpha_s \cos pr_s}{pr_s(1+ip\alpha_s)}.$$

## 3 Short-distance P-odd interaction

We are interested first of all in the close vicinity of the threshold where the deuteron disintegration is dominated by the regular $M1$ transition $^3S_1 \to {}^1S_0$. In the admixed $E1$ transition the total spin is conserved. Therefore here we need the P-odd weak interaction which does not conserve the total spin (and conserves the isotopic spin). This interaction admixes $^1P_1$ state to the initial one $^3S_1$, and $^3P_0$ to the final state $^1S_0$. This interaction is of a short-range nature, and we will use its common description by a potential, corresponding to the $\rho$-, $\omega$-exchange. Due to the shortcomings of this description, pointed out in Introduction, quantitative results

| $g_\rho$ | $g_\omega$ | $\chi_\rho$ | $\chi_\omega$ | $h_\rho^0 \times 10^7$ | $h_\rho^1 \times 10^7$ | $h_\rho^2 \times 10^7$ | $h_\omega^0 \times 10^7$ | $h_\omega^1 \times 10^7$ |
|---|---|---|---|---|---|---|---|---|
| 2.79 | 8.37 | 3.7 | $-0.12$ | $-11.4$ | 0 | $-9.5$ | $-1.9$ | 0 |

Table 1: Numerical values of the constants in potential (9)

obtained here should be considered as detailed estimates only. The short-range P-odd potential is [28]

$$W = -g_\rho \left[ h_\rho^0 \boldsymbol{\tau}_1 \boldsymbol{\tau}_2 + \frac{1}{2} h_\rho^1 (\tau_1^z + \tau_2^z) + \frac{1}{2\sqrt{6}} h_\rho^2 (3\tau_1^z \tau_2^z - \boldsymbol{\tau}_1 \boldsymbol{\tau}_2) \right]$$
$$\times \frac{1}{2m_p} \left( (\boldsymbol{\sigma}_1 - \boldsymbol{\sigma}_2)\{\mathbf{p}_1 - \mathbf{p}_2, f_\rho(r)\} + 2(1 + \chi_\rho)[\boldsymbol{\sigma}_1 \times \boldsymbol{\sigma}_2] \boldsymbol{\nabla} f_\rho(r) \right)$$
$$- g_\omega \left[ h_\omega^0 \boldsymbol{\tau}_1 \boldsymbol{\tau}_2 + \frac{1}{2} h_\omega^1 (\tau_1^z + \tau_2^z) \right] \qquad (9)$$
$$\times \frac{1}{2m_p} \left( (\boldsymbol{\sigma}_1 - \boldsymbol{\sigma}_2)\{\mathbf{p}_1 - \mathbf{p}_2, f_\omega(r)\} + 2(1 + \chi_\omega)[\boldsymbol{\sigma}_1 \times \boldsymbol{\sigma}_2] \boldsymbol{\nabla} f_\omega(r) \right)$$
$$- \frac{1}{2} (\tau_1^z - \tau_2^z)(\boldsymbol{\sigma}_1 + \boldsymbol{\sigma}_2) \frac{1}{2m_p} \{\mathbf{p}_1 - \mathbf{p}_2, g_\omega h_\omega^1 f_\omega(r) - g_\rho h_\rho^1 f_\rho(r)\}.$$

Here
$$f_{\rho,\omega}(r) = \frac{e^{-m_{\rho,\omega} r}}{4\pi r}, \quad r = |\mathbf{r}_1 - \mathbf{r}_2|.$$

The numerical values of the parameters entering this expression are presented in Table 1. The values of the constants corresponding to strong vertices, $g_{\rho,\omega}$, $\chi_{\rho,\omega}$, are reasonably reliable. For the P-odd constants, $h_\rho^{0,1,2}$, $h_\omega^{0,1}$, we use the "best values" [7].

The last term in (9), which conserves the total spin $\mathbf{I} = (\boldsymbol{\sigma}_p + \boldsymbol{\sigma}_n)/2$, will be considered below, together with the P-odd pion exchange. Here we treat other terms in (9) which do not conserve $\mathbf{I}$ (but conserve the isotopic spin).

Let us start with the correction to the deuteron wave function $\psi_d$, using the common stationary perturbation theory. In the ZRA approximation the admixed $P$ states of the continuous spectrum are free. Moreover, we can choose plane waves as the intermediate states since the perturbation (9) selects by itself the $P$ state from the plane wave. Thus obtained correction can be written as

$$\delta\psi_d(\mathbf{r}) = \int \frac{d\mathbf{k}}{(2\pi)^3} \frac{e^{i\mathbf{kr}}}{-\varepsilon - k^2/m_p} \int d\mathbf{r}' e^{-i\mathbf{kr}'} W(\mathbf{r}')\psi_d(r') \chi_t$$
$$= -\frac{m_p}{4\pi} \int d\mathbf{r}' \frac{e^{-\kappa|r-r'|}}{|r-r'|} W(r')\psi_d(r') \chi_t \qquad (10)$$
$$= \frac{m_p}{4\pi} \boldsymbol{\nabla} \frac{e^{-\kappa r}}{r} \int d\mathbf{r}' r' W(r')\psi_d(r') \chi_t,$$

the last transformation being possible due to the short-range nature of $W(\mathbf{r})$ ($\kappa \ll m_{\rho,\omega}$). Here $\chi_t$ is the triplet spin wave function of the deuteron (previously we omitted it for brevity). Simple algebra transforms this expression into

$$\delta\psi_d = -i\lambda_t (\mathbf{\Sigma}\mathbf{\nabla}) \sqrt{\frac{\kappa}{2\pi}} \frac{e^{-\kappa r}}{r} \chi_t, \tag{11}$$

where $\mathbf{\Sigma} = \boldsymbol{\sigma}_p - \boldsymbol{\sigma}_n$. Though this P-odd admixture of the $^1P_1$ state to $^3S_1$ is expressed conveniently through the ZRA wave function, the constant $\lambda_t$ introduced in (11), depends in fact on the true deuteron wave function $\psi_d$ as follows:

$$\sqrt{\frac{\kappa}{2\pi}} \lambda_t \chi_t = -i \frac{m_p}{16\pi} \int d\mathbf{r}'(\mathbf{\Sigma}\mathbf{r}')W(\mathbf{r}')\psi_d(r')\,\chi_t\,.$$

In the same way we calculate the P-odd $^3P_0$ admixture to the wave function of the $^1S_0$ state of continuous spectrum:

$$\delta\psi_{Ss} = i\lambda_s a_s (\mathbf{\Sigma}\mathbf{\nabla}) \frac{e^{ipr}}{r} \chi_s\,. \tag{12}$$

Here $\chi_s$ is the singlet spin wave function, and the constant $\lambda_s$ is expressed via the wave function $\psi_{Ss}$ of the $^1S_0$ state (see (8)) as follows:

$$\lambda_s \chi_s = -i \frac{m_p}{48\pi} \int d\mathbf{r}'(\mathbf{\Sigma}\mathbf{r}')W(\mathbf{r}')\psi_{Ss}(r')\,\chi_t\,.$$

We will need also the P-odd $^1S_0$ admixture to the $^3P_0$ state of continuous spectrum. It can be easily found from the requirement that the perturbed $^1S_0$ and $^3P_0$ wave functions should remain orthogonal. We obtain

$$\delta\psi_P = -2i\lambda_s a_s p \frac{e^{ipr}}{r} \chi_s. \tag{13}$$

A general formula comprising all three cases, (11), (12), (13), was given previously in [12]. Straightforward calculations with wave functions (11) and (12) give

$$\sigma_+ - \sigma_- = -\frac{16\pi\alpha}{9} \frac{\kappa p\,(\mu_p - \mu_n)(1 - \kappa\alpha_s)(3\kappa^2 + p^2)}{m_p(\kappa^2 + p^2)^2(1 + p^2\alpha_s^2)}$$

$$\times \left[\lambda_s \kappa \alpha_s \frac{\kappa^2 + 3p^2}{3\kappa^2 + p^2} + \lambda_t \left(1 - \kappa\alpha_s \frac{2\kappa^2}{3\kappa^2 + p^2}\right)\right]. \tag{14}$$

An analogous formula for the deuteron disintegration by longitudinally polarized electrons was derived in [19].

Wave functions (11) and (13) induce a P-odd asymmetry in one more way. Here the regular amplitude is $E1$: $^3S_1 \to {}^3P_{0,1,2}$ and the admixed amplitudes are $M1$: $^1P_1 \to {}^3P_{0,1,2}$ and $^3S_1 \to {}^1S_0$. This contribution to the P-odd cross-section difference is

$$\sigma_+ - \sigma_- = -\frac{32\pi\alpha}{9} \frac{\kappa p^3\,(\mu_p - \mu_n)}{m_p(\kappa^2 + p^2)^2\sqrt{1 - \kappa r_t}} \left[\lambda_s \kappa \alpha_s \frac{1 + p^2\alpha_s/\kappa}{1 + p^2\alpha_s^2} - \lambda_t\right]. \tag{15}$$

While the above derivation of expressions (14), (15) is a relatively simple procedure, the problem of calculating the constants $\lambda_s$ and $\lambda_t$ is quite different. In the ZRA the calculation of these constants is rather straightforward and results in

$$\lambda_s^{(zra)} = (0.153 h_\rho^0 - 0.124 h_\rho^2 + 0.229 h_\omega^0) \times 10^{-7} m_\pi^{-1} = -1.09 \times 10^{-7} m_\pi^{-1},$$

$$\lambda_t^{(zra)} = (0.281 h_\rho^0 + 0.116 h_\omega^0) \times 10^{-7} m_\pi^{-1} = -4.52 \times 10^{-7} m_\pi^{-1}. \tag{16}$$

However, these naïve ZRA numbers for the effective constants $\lambda_{s,t}$ certainly strongly overestimate true values of these constants. The first reason is that the ZRA wave functions of the $S$-states are singular at $r \to 0$, while their correct wave functions are finite at the origin. Therefore, here we will use instead of naïve ZRA wave functions (1) and (2), model functions (7) and (8) which are finite at the origin.

By the same reason of the short-range nature of vector exchanges, one more suppression factor is essential here. We mean the Jastrow repulsion between nucleons at small distances. Following [29-31], we will take it into account by a factor $\phi^2(r)$ in the weak matrix elements, where

$$\phi(r) = 1 - c e^{-dr^2}, \quad c = 0.6, \quad d = 3 \text{ fm}^{-2}. \tag{17}$$

With these modifications, we arrive at more realistic (and much smaller!) estimates for the constants $\lambda_s$ and $\lambda_t$:

$$\lambda_s = (0.028 h_\rho^0 - 0.023 h_\rho^2 + 0.028 h_\omega^0) \times 10^{-7} m_\pi^{-1} = -0.16 \times 10^{-7} m_\pi^{-1},$$

$$\lambda_t = (0.032 h_\rho^0 + 0.001 h_\omega^0) \times 10^{-7} m_\pi^{-1} = -0.37 \times 10^{-7} m_\pi^{-1}. \tag{18}$$

The P-odd asymmetry

$$A = \frac{\sigma_+ - \sigma_-}{\sigma_+ + \sigma_-} \tag{19}$$

of the deuteron photodisintegration due to the cross-section differences (14) and (15), as calculated with the constants (18), is plotted in Figs. 1a,b, respectively. We have chosen different vertical scales in Figs. 1a,b to be able to reproduce details of the effects differing by two orders of magnitude. Obviously, in the whole range of energies considered, the contribution corresponding to the regular $E1$ transition is very small, and can be safely neglected.

## 4 Spin-conserving P-odd interaction

It was mentioned already that $M1$ transition from the ground state proceeds only to the $^1S_0$ state of the continuous spectrum. Then, it can be easily seen that the P-odd exchange, which conserves the total spin $\mathbf{I}$, operates in our problem as follows. In the regular $E1$ transition from the ground state $^3S_1$ into $^3P_{0,1,2}$, it admixes $^3P_1$ state of the continuous spectrum to the initial one, and $^3S_1$ state to the final $^3P_1$ one. To contribute to the admixed P-odd $M1$ amplitude, this last admixed $^3S_1$ state should be the ground state of the deuteron (recall the mentioned orthogonality of the $^3S_1$ radial wave functions of different energies). With the account for this P-odd mixing the mentioned states of $J = 1$ can be written as

$$^3\tilde{S}_1 = {}^3S_1 + i\beta \, {}^3P_1, \quad {}^3\tilde{P}_1 = {}^3P_1 + i\beta \, {}^3S_1,$$

where $i\beta$ is purely imaginary weak mixing coefficient. Then straightforward standard calculations with $3j$ and $6j$ symbols demonstrate that the sums of the reduced $E1$ and $M1$ amplitudes for each of the electromagnetic transitions $^3\tilde{S}_1 \to {}^3\tilde{P}_0, {}^3\tilde{P}_1, {}^3P_2$ look as follows (up to a factor common to all three transitions):

$$^3\tilde{S}_1 \longrightarrow \begin{cases} {}^3P_0 & 1 - \lambda\beta\rho \\ {}^3\tilde{P}_1 & -\sqrt{3}\left(1 - \tfrac{1}{2}\lambda\beta\rho\right) \\ {}^3P_2 & \sqrt{5}\left(1 + \tfrac{1}{2}\lambda\beta\rho\right) \end{cases} \tag{20}$$

Here $\lambda = \pm$ is the sign of the photon circular polarization, and $\rho$ is the ratio of the reduced $E1$ and $M1$ amplitudes. Now, the total probability of the $\gamma d \to np$ reaction is proportional to

$$1 - \lambda\beta\rho + 3(1 - \lambda\beta\rho) + 5(1 + \lambda\beta\rho),$$

and obviously independent of the circular polarization $\lambda$. Thus, the P-odd weak interaction which conserves the total spin does not contribute at all to the discussed asymmetry in the deuteron photodisintegration. It refers both to the weak pion exchange[2] and to the corresponding part of the short-distance contribution (the last line in formula (9)).

## 5  Conclusions

Our final result for the total asymmetry $A$ practically coincides with the curve plotted in Fig. 1a, in the whole region of the energies discussed (the second of the short-distance contributions, plotted in Fig. 1b, is negligibly small, as compared to the first one, for these energies). The maximum value of the asymmetry, at the threshold, is about $10^{-7}$.

Unfortunately, the magnitude of the short-distance effect cannot be accurately predicted. Our result for it is higher, by a factor of 2 to 5, than those of previous works, which also differ considerably among themselves. Different approaches and lack of details of calculations in those papers preclude elucidation of the exact origin of this disagreement. However, at least in the case of our discrepancy with [16] there is a plausible explanation for it. The cut-off adopted in [16] for the description of the short-range nucleon-nucleon repulsion (the short-range correlation factor therein turns to zero for $r < r_c$, $r_c = 0.43$ fm or $0.56$ fm) is much more steep than the cut-off adopted by us (see (17)). The argument in favor of our approach was presented already in Introduction. It is the good agreement between the experimental value of the anapole moment of $^{133}Cs$ and the theoretical prediction for it obtained within the approach used here.

Our last remark refers to the relation between the P-odd asymmetry $A$ and the degree of circular polarization $P$ of $\gamma$-quanta in the inverse reaction $np \to d\gamma$:

$$P = \frac{\tilde{\sigma}_+ - \tilde{\sigma}_-}{\tilde{\sigma}_+ + \tilde{\sigma}_-}. \qquad (21)$$

In this expression $\tilde{\sigma}_\lambda$ is the production cross-section for a photon with circular polarization $\lambda(= \pm)$. In virtue of the principle of detailed balancing (which is valid here since the interactions considered are T-even),

$$A = P. \qquad (22)$$

If our threshold value for $A$ is correct, i.e., if indeed

$$A \approx 1.0 \times 10^{-7},$$

then, according to (22), the experimental upper limit for $P$ obtained in [1],

$$P = (1.8 \pm 1.8) \times 10^{-7},$$

is close to the real effect.

---

[2] A nonvanishing value for the pion exchange contribution to the asymmetry was obtained in [24] as a result of the incomplete account for P-odd mixing: only the $^3P_1$ admixture to the deuteron ground state was included there.

∗∗∗

We very much appreciate stimulating discussions with G.Ya. Kezerashvili (deceased), G.N. Kulipanov, D.M. Nikolenko, and D.K. Toporkov, our interest to the problem is essentially due to them. We are extremely grateful to B. Desplanques for criticisms, and in particular, for pointing out to us the possibility that the pion contribution to the discussed asymmetry vanishes. The work was supported by the Russian Foundation for Basic Research through Grant No. 98-02-17797, by the Ministry of Education through Grant No. 3N-224-98, and by the Federal Program Integration-1998 through Project No. 274.